\documentclass[aps, prd, twocolumn, lengthcheck, superscriptaddress, showpacs, letterpaper, nofootinbib]{revtex4-1}

\usepackage{amsfonts}
\usepackage{amsmath}
\usepackage{amssymb}
\usepackage{xcolor}
\usepackage{graphicx}
\graphicspath{{figures/}}
\usepackage{subfigure}
\usepackage{longtable}
\usepackage{slashed}
\usepackage{hyperref}
\usepackage{amsthm}
\usepackage{mathrsfs}
\usepackage{color}
\usepackage{epsfig}
\usepackage{epstopdf}

\def\be{\begin{eqnarray}}\def\ee{\end{eqnarray}}

\def\be{\begin{eqnarray}}\def\ee{\end{eqnarray}}
\def\lsim{\mathrel{\rlap{\lower3pt\hbox{\hskip1pt$\sim$}}
     \raise1pt\hbox{$<$}}} 
\def\gsim{\mathrel{\rlap{\lower3pt\hbox{\hskip1pt$\sim$}}
     \raise1pt\hbox{$>$}}} 

\allowdisplaybreaks

\begin{document}

\title{Scale symmetry and composition of compact star matter}

\author{Long-Qi Shao}
\affiliation{College of Physics, Jilin University, Changchun 130012, China}
\affiliation{School of Fundamental Physics and Mathematical Sciences, Hangzhou Institute for Advanced Study, UCAS, Hangzhou, 310024, China}

\author{Yong-Liang Ma}
\email{ylma@ucas.ac.cn}
\affiliation{School of Fundamental Physics and Mathematical Sciences, Hangzhou Institute for Advanced Study, UCAS, Hangzhou, 310024, China}
\affiliation{College of Physics, Jilin University, Changchun 130012, China}

\begin{abstract}
The dense compact star matter is studied by using the skyrmion crystal approach. The chiral effective theory used includes the lightest scalar meson, the lowest-lying vector mesons as well as pions. Consistency with the vector manifestation and the dilaton limit fixed point at high density constrains the anomalous dimension of the gluon field $1.0 \lesssim |\gamma_{G^2}| \lesssim 2.0$ and leads to the significance of the scale symmetry breaking in the intrinsic parity-odd part of the effective theory. The speed of sound $v_s^2 \simeq 1/3$ and the polytropic index $\gamma \simeq 1$---both satisfy the conformal limits---after the dilaton limit fixed point at high density but the matter is still in the hadronic phase. This means that neither the conformal speed of sound nor the smallness of the polytropic index can be used as a criterion of the onset of quark matter. These conclusions are significant for constructing the equation of state of nuclear matter.
\end{abstract}


\maketitle

\section{introduction}

Dense nuclear matter relevant to the core of compact star is an uncharted domain although it has been studied for several decades~\cite{Brown:2001nh,Li:2008gp,Holt:2014hma,Drews:2016wpi,Li:2019xxz,Ma:2019ery,Kim:2020mhk,Ma:2020nih}. Several fundamental but significant questions in dense nuclear matter, such as the pattern of symmetry, the constituents and so on are not yet well understood. The recent observations of massive neutron stars with about two times solar mass and detections of the gravitational waves from binary neutron star mergers and neutron star-black hole mergers open new laboratories for dense nuclear matter (see, e.g., ~\cite{Li:2019xxz,Ma:2019ery,Kim:2020mhk,Ma:2020nih} and references therein).

The access to dense nuclear matter is mainly based on the models or effective theories anchored on the symmetries of QCD. The parameters are fixed by using the nuclear matter properties around saturation density. The properties of dense nuclear matter are predicted by extrapolating such obtained equation of state (EoS)~(e.g., Refs.~\cite{Drews:2016wpi,Furnstahl:2019lue}). Whether this extrapolation is reasonable is questionable. An alternative access is to implement the conjectured symmetry/constituent to the construction of EoS (e.g.~\cite{Alford:2017qgh}). This construction is highly model dependent. Therefore, it is valuable to obtain some information of dense nuclear matter using a unified model.

It is widely accepted that when considered at the large $N_c$ limit, baryons can be described as skyrmions~\cite{Skyrme:1961vq,Witten:1979kh} and consequently skyrmion matter can be regarded as nuclear matter~\cite{Klebanov:1985qi}. The skyrmion approach provides a model independent way to study nucleon, nuclei and nuclear matter in a unified model in the sense of large $N_c$ limit~\cite{Zahed:1986qz,Naya:2018kyi,Ma:2016gdd}. As a result, it is interesting to use the skyrmion approach to extract some model independent information of dense nuclear matter where the large $N_c$ argument is applicable~\cite{McLerran:2007qj,McLerran:2018hbz,Zhao:2020dvu}.

Since the isoscalar scalar meson sigma and isoscalar vector meson omega are indispensable ingredients for the nuclear force~\cite{Zeldovich:1961sbr,Serot:1984ey}, we include both in the calculation. The scalar meson is introduced as the dilaton $\chi$, the Nambu-Goldstone boson of scale symmetry breaking~\cite{Crewther:2013vea,Crewther:2020tgd}. The omega meson and its flavor partners rho mesons are included through the hidden local flavor symmetry~\cite{Bando:1984ej,Bando:1987br,Harada:2003jx}. Both symmetries are hidden in the matter free space and are expected to emerge in the (super-)highly condensed matter. In summary, the effective theory (denoted as $s$HLS) used in this work includes the lightest scalar meson, omega, rho and the pseudo-Nambu Goldstone bosons pions.

Based on the Wilsonian renormalization group (RG) approach, it is found that at high energy scale $f_\pi^\ast \to 0, m_\rho^\ast \to m_\pi \to 0$, i.e., there is a vector manifestation (VM) fixed point in the hidden local symmetry (HLS)~\cite{Harada:2000kb,Harada:2003jx}. It is not strange to expect that the VM appears at (super-)high density. In addition, in the approach to the baryonic matter using the dilaton compensated chiral effective theory, people found that there is a dilaton limit (DL) fixed point which states that the medium modified decay constant $f_\sigma^\ast \to 0$ in the theory at high density~\cite{Beane:1994ds,Paeng:2011hy}. We expect both the VM fixed point and the DL fixed point to be realized in the skyrmion crystal approach.

The purpose of this work is to study how the scale symmetry is realized in dense nuclear matter and what is the possible constituents in the cores of the massive neutron stars using skyrmion crystal approach based on $s$HLS. The $s$HLS to be specified later includes the leading order of the chiral-scale counting, the intrinsic parity-odd terms as well as a dilaton potential which breaks the scale symmetry both explicitly and spontaneously. Although the homogeneous Wess-Zumino (hWZ) term, the intrinsic parity-odd term is scale invariant, we couple dilaton to it to obtain the finite energy of the skyrmion matter~\cite{Park:2008zg}. Therefore, the dilaton compensated hWZ term does not conserve the scale symmetry. Different from~\cite{Park:2008zg} where the dilaton couples to the hWZ term in an arbitrary way, the coupling between dilaton and hWZ term has a solid physical argument in the genuine dilaton approach~\cite{Crewther:2013vea,Crewther:2020tgd}.

We found that the requirement $2\lesssim n_{\rm DLFP}/n_0 \lesssim 10$ with $n_0$ being the saturation density of nuclear matter and $n_{\rm DLFP}$ being the onset of the DL fixed point (the same as the VM fixed point in the present approach) yields the magnitude of the anomalous dimension of gluon field $1.0 \lesssim |\gamma_{G^2}| \lesssim 2.0$ which is consistent with~\cite{Park:2008zg,Ma:2016nki}. More interestingly, we found that after $n_{\rm DLFP}$, the speed of sound $v_s^2 \to 1/3$ and the polytropic index $\gamma \to 1$ at density relevant to the cores of massive neutron stars, both satisfy the constraints from conformal invariance. This indicates that the speed of sound of the compact star matter could be $v_s^2 \approx 1/3$~\cite{Ma:2018qkg,Zhao:2020dvu,Kapusta:2021ney,Margueron:2021dtx}, in contrast to the usual conclusion in the literature \cite{Bedaque:2014sqa,Tews:2018kmu,Moustakidis:2016sab}. Note that the conformal speed of sound of the dense matter after $n_{\rm DLFP}$ does not mean that the matter is scale invariant since the trace of the energy-momentum tensor (TEMT) is not zero but a density independent constant. Therefore, the matter after $n_{\rm DLFP}$ is pseudoconformal~\cite{Paeng:2017qvp,Ma:2018xjw,Ma:2018jze,Ma:2019ery}. In contrast to the usual picture of dense nuclear matter in which the conformal speed of sound $v_s^2 =1$ and polytropic index $\gamma \to 1$ are regarded as the criterions of the emergence of the deconfined quark~\cite{Annala:2019puf}, there is no deconfined quark in the core of massive stars in this pseudoconformal nuclear matter~\cite{Ma:2020hno}.

We organize this paper as follows: In Sec.~\ref{sec:EFT} we present the key points of the $s$HLS used in this work. In Sec.~\ref{sec:property} we discuss the nuclear matter properties calculated from the skyrmion crystal in detail. Our conclusions and discussions are presented in Sec.~\ref{sec:Conclusion}.

\section{Hidden scale and Hidden local symmetric Lagrangian}

\label{sec:EFT}

The Skyrme type model we are going to consider in the following is based on the hidden local symmetry (HLS)~\cite{Bando:1984ej,Bando:1987br,Harada:2003jx}. Among a variety of approaches to the model of the lightest scalar meson~\cite{Crewther:2013vea,Golterman:2016lsd,Crewther:2020tgd,DelDebbio:2021xwu}, we here use the genuine dilaton approach~\cite{Crewther:2013vea,Crewther:2020tgd} in a minimal way such that both the chiral and scale symmetries can (partially) restore in matter.

The dilaton compensated HLS (denoted as $s$HLS) can be written as \cite{Crewther:2013vea,Li:2016uzn}
\be
\mathcal{L}_{s \mathrm{HLS}} & = & \mathcal{L}_{(2)}^{s \mathrm{HLS}}+\tilde{\mathcal{L}}_{\mathrm{hWZ}}+\mathcal{V}(\chi). \label{eq:LagFull}
\ee
The leading chiral-scale counting order Lagrangian $\mathcal{L}_{(2)}^{s \mathrm{HLS}}$ is
\be
\mathcal{L}_{(2)}^{s \mathrm{HLS}} & = & f_{\pi}^{2}\left(\frac{\chi}{f_{\sigma}}\right)^{2} {\rm Tr}\left[\hat{a}_{\perp \mu} \hat{a}_{\perp}^{\mu}\right]+a f_{\pi}^{2}\left(\frac{\chi}{f_{\sigma}}\right)^{2} {\rm Tr}\left[\hat{a}_{\| \mu} \hat{a}_{\|}^{\mu}\right] \nonumber\\
& & {} -\frac{1}{2 g^{2}} \operatorname{Tr}\left[V_{\mu \nu} V^{\mu \nu}\right]+\frac{1}{2} \partial_{\mu} \chi \partial^{\mu} \chi      \label{eq:Op2}
\ee
where the pion fields are $U(x)=\exp(2i\pi/f_\pi)=\xi_L^\dagger \xi_R$ and
\be
\hat{a}_{\perp \mu} & = & \frac{1}{2i}\left(D_\mu \xi_R\cdot \xi_R^\dagger - D_\mu \xi_L\cdot \xi_L^\dagger \right), \nonumber\\
\hat{a}_{\parallel \mu} & = & \frac{1}{2i}\left(D_\mu \xi_R\cdot \xi_R^\dagger + D_\mu \xi_L\cdot \xi_L^\dagger \right).
\ee
The covariant derivative $D_\mu \xi_{L,R} = \partial_\mu \xi_{L,R} - iV_\mu \xi_{L,R}$ with $V_\mu$ being the gauge field of the HLS. In this work, we take hidden local symmetry as $H_{\rm local} = U(2)$ and identify
\be
V_\mu = g\left(\rho_\mu + \omega_\mu\right),
\ee
where $g$ is the gauge coupling constant of HLS. In Eq.~\eqref{eq:Op2}, the conformal compensator field $\chi(x)=f_\sigma e^{\sigma/f_\sigma}$ with $\sigma$ being the dilaton field and $f_\sigma$ being the decay constant of dilaton. Through out this work, we take the chiral limit.

The potential $\mathcal{V}(x)$ in our approach is responsible for both the spontaneous and explicit breaking of scale symmetry. After taking the saddle point equation we can write the dilaton potential in terms of dilaton mass as
\be
V_{\chi}={}-\frac{f_{\sigma}^{2} m_{\sigma}^{2}}{4 \beta^{\prime}} \left(\frac{\chi}{f_\sigma}\right)^4+\frac{f_{\sigma}^{2} m_{\sigma}^{2}}{\beta^{\prime}\left(4+\beta^{\prime}\right)} \left(\frac{\chi}{f_\sigma}\right)^{4+\beta^{\prime}},
\label{eq:DPotential}
\ee
where $\beta^\prime = |\gamma_{G^2}|$ is the nonperturbative anomalous dimension of gluon field which cannot be calculated from QCD. Note that the dilaton potential \eqref{eq:DPotential} is different from the logarithm form used in~\cite{Ma:2016nki}. The logarithm form in ~\cite{Ma:2016nki} can be reduced to from \eqref{eq:DPotential} by taking $\beta^\prime \ll 1$ so than Eq.~\eqref{eq:DPotential} is a more general form of the potential. We will find that the general expression of the dilaton potential \eqref{eq:DPotential} leads to a more stringent constraint on the nonperturbative anomalous dimension of gluon field.

How the scale symmetry breaking effect is involved in the intrinsic parity-odd part $\mathcal{L}_{\rm hWZ}$ is a subtle question. Here, for simplicity and without lose of generality, we take a universal scale symmetry breaking form
\be	
\tilde{\mathcal{L}}_{\mathrm{hWZ}} = \left(c_{h}+\left(1-c_{h}\right)\left(\frac{\chi}{f_{\sigma}}\right)^{\beta^{\prime}}\right) \mathcal{L}_{\mathrm{hWZ}}, \label{eq:hWZ}
\ee
where $c_h$ is a free parameter. When $c_h=1$ or $\langle \chi \rangle = f_\sigma$, the dilaton compensator disappears and $\tilde{\mathcal{L}}_{\rm hWZ}$ reduces to the scale invariant form $\mathcal{L}_{\rm hWZ}$. The hWZ term $\mathcal{L}_{\rm hWZ}$ takes the form
\be
\mathcal{L}_{\mathrm{hWZ}} & = & \frac{N_{c}}{16 \pi^{2}} \sum_{i=1}^{3} c_{i} \mathcal{L}_{i}  \label{eq:GammahWZ}
\ee
where in terms of the 1-form and 2-form notations
\begin{subequations}
\be
\mathcal{L}_{1} & = & i \operatorname{Tr}\left[\hat{\alpha}_{\mathrm{L}}^{3} \hat{\alpha}_{\mathrm{R}}-\hat{\alpha}_{\mathrm{R}}^{3} \hat{\alpha}_{\mathrm{L}}\right], \label{6a}\\
\mathcal{L}_{2} & = & i \operatorname{Tr}\left[\hat{\alpha}_{\mathrm{L}} \hat{\alpha}_{\mathrm{R}} \hat{\alpha}_{\mathrm{L}} \hat{\alpha}_{\mathrm{R}}\right],\label{6b}\\
\mathcal{L}_{3} & = & \operatorname{Tr}\left[F_{\mathrm{V}}\left(\hat{\alpha}_{\mathrm{L}} \hat{\alpha}_{\mathrm{R}}-\hat{\alpha}_{\mathrm{R}} \hat{\alpha}_{\mathrm{L}}\right)\right],\label{6c}
\ee
\end{subequations}
with
\begin{align}
\hat{\alpha}_{L} & = \hat{\alpha}_{\|}-\hat{\alpha}_{\perp},\notag\\
\hat{\alpha}_{R} & = \hat{\alpha}_{\|}+\hat{\alpha}_{\perp},\notag\\
F_{V} & = d V-i V^{2}. \label{7}
\end{align}
Since in the medium $\left\langle \chi\right\rangle^\ast /f_\sigma<1$, when $c_h\ne1$ and $\beta^\prime\neq 0$, the scale symmetry breaking is included in the intrinsic parity-odd part. In this note, we take $c_h$ as well as the anomalous dimension $\beta^\prime$ as unknown parameters and fix them by requiring the consistency to the VM and DL fixed points. It should be noted that when $c_h = 0$ and $\beta=3$, a specific choice of the low energy constants in HLS, one reduces to the model of Ref.~\cite{Park:2008zg}.

\section{ Structure of compact star matter}

\label{sec:property}

Now we are ready to calculate the nuclear matter by using the skyrmion crystal approach. This is accessed by putting the skyrmions calculated from Lagrangian \eqref{eq:LagFull} to a specific crystal lattice~\cite{Klebanov:1985qi}. Here we take the face-centered cubic (FCC) lattice which is known yields the lowest ground state energy so far~\cite{KUGLER1988491}.

In the numerical calculation, we take the typical values $m_\sigma= 640$~MeV and $f_\sigma=240$~MeV and run $\beta^\prime \in (0,4)$ and $c_h \in (0,1)$. The low energy constants $c_i$ cannot be determined phenomenologically so far. Here we choose $c_1 = - c_2 = 2/3$ and $c_3 =0$ such that
\be	
\tilde{\mathcal{L}}_{\mathrm{hWZ}} = g \omega_\mu B^\mu \left(c_{h}+\left(1-c_{h}\right)\left(\frac{\chi}{f_{\sigma}}\right)^{\beta^{\prime}}\right),
\ee
with $B_\mu$ being the topology (baryon) current---the minimal model~\cite{Meissner:1986js}. To illustrate the parameter independence of the existence of DL fixed point, we compare the results with another set of values $c_1 = 2/3, c_2 = -1/3$ and $c_3=1$ which are fixed by the possible Seiberg duality that identifies the $\omega$ field with the Chern-Simons field together with the consistence of vector meson dominance~\cite{Karasik:2020zyo}.

\subsection{Dilaton limit fixed point and vector manifestation}

We first study the constraints on $c_h$ and $\beta^\prime$ by requiring the consistency to the VM/DL fixed point, typically, at density $2\lesssim n_{\rm DLFP}/n_0 \lesssim 10$. From Lagrangian~\eqref{eq:LagFull}, one can obtain
\be
f_\pi^\ast \propto m_\rho^\ast \propto f_\sigma^\ast.
\ee
Therefore, the VM fixed point is locked to the DL fixed point. Note that since the general dilaton potential~\eqref{eq:DPotential} is different from that used in~\cite{Ma:2016nki} which is valid only $\beta^\prime \ll 1$, the constraints are found to be $0 \lesssim c_h \lesssim 0.2$ and $1.0\lesssim \beta^\prime \lesssim 2.0$, consistent to but slightly different from that obtained in ~\cite{Ma:2016nki}.

\begin{figure}[htp]
	\includegraphics[width=9cm]{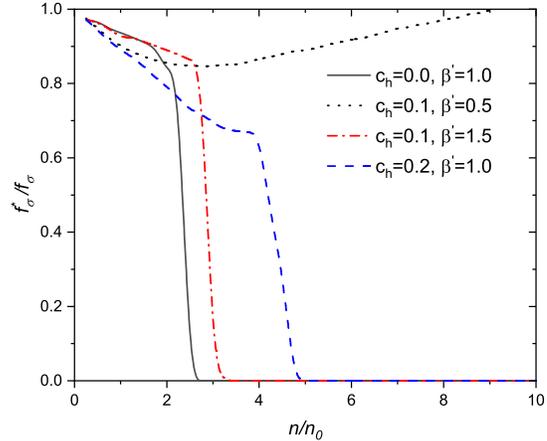}
	\caption{Density dependence of the medium modified dilaton decay constant $f_\sigma^\ast$.}
	\label{fig:fsigma}
\end{figure}

The density dependence of $f_\sigma^\ast/f_\sigma$ is shown in Fig.~\ref{fig:fsigma} with some typical values of $c_h$ and $\beta^\prime$. One can easily see that, for a fixed $\beta^\prime$, the bigger $c_h$, the larger $n_{\rm DLFP}$. This is because, the bigger $c_h$, the less omega repulsive force is affected by density and the less omega force is screened therefore the more difficult $n_{\rm DLFP}$ can be arrived. When $c_h$ is bigger than a certain typical value---$\approx 0.2$---the omega force is not screened enough so that the scale restoration cannot happen. In addition, for a fixed $c_h$, the smaller $\beta^\prime$ the larger $n_{\rm DLFP}$ because of the larger repulsive force from omega meson. When $\beta^\prime$ is smaller than a typical value---$\approx 0.5$---the DL/VM fixed point cannot appear and when $\beta^\prime$ exceeds a typical value---$\approx 2.0$---the DL/VM fixed point appears at a low density that inconsistent with nature.

The result from both $c_h \neq 1$ and $\beta^\prime \neq0$ implies the significance of the scale symmetry breaking in the hWZ part~\cite{Park:2008zg,Ma:2016nki}. Moreover, it tells us that in between the low density---which cannot be accessed in the present approach---and the DL fixed point, the symmetry is not visible, hence hidden~\cite{Ma:2020tsj}.

\subsection{Pseudocomformal dense nuclear matter}

\begin{figure}[htp]
	\includegraphics[width=8cm]{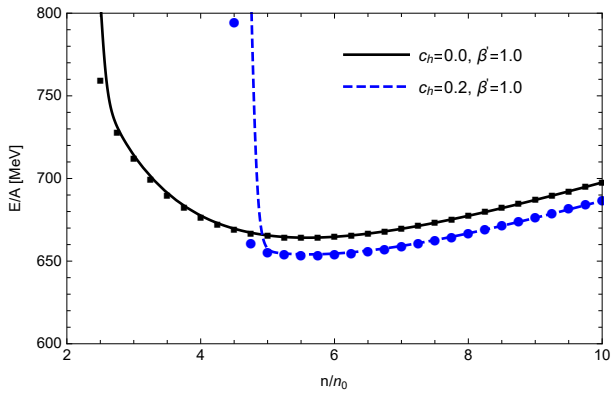}
	\caption{Comparison of the crystal data and fitted curves.}
	\label{fig:EAFit}
\end{figure}

In the skyrmion crystal approach, the per-skyrmion energy $E/A$ in the medium can be calculated at a fixed crystal size, or equivalently, density~\cite{Klebanov:1985qi}. Interestingly we found that when $n \gtrsim n_{\rm DLFP} $, $E/A$ can be nicely fitted by the following function
\be
E/A = a \left(\frac{n}{n_0}\right)^{1/3} + b \left(\frac{n}{n_0}\right)^{-1}.
\label{eq:FitFunction}
\ee
The fitted curves compared to the crystal data points are shown in Fig.~\ref{fig:EAFit}.

From function \eqref{eq:FitFunction} one can easily conclude that the speed of sound satisfies the conformal limit
\be
v_s^2 & = & \frac{\partial P(n)}{\partial n}/\frac{\partial\epsilon(n)}{\partial n} \to 1/3,
\ee
where $P(n)$ and $\epsilon(n)$ are, respectively, the pressure density and energy density. In addition, one finds that the polytropic index $\gamma$ in the nuclear matter approaches to
\be
\gamma = \frac{d\ln P}{d\ln \epsilon} \to 1,
\ee
when $n \gtrsim n_{\rm DLFP}$. It also satisfies the constraint from conformal invariance.

 \begin{figure}[htp]
	\includegraphics[width=8cm, height=6cm]{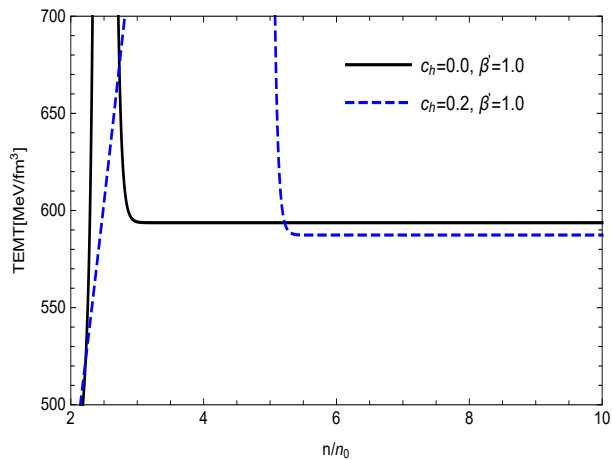}
	\caption{Trace of energy-momentum tensor as a function of density.}
	\label{fig:TEMT}
 \end{figure}

How to understand $v_s^2 \to 1/3$ in the dense matter? Does it mean that the matter becomes a conformal invariant theory at high density? To have a deeper understanding of the structure of the dense nuclear matter, let us check the density dependence of the trace of the energy-momentum tensor (TEMT) using the fitting function~\eqref{eq:FitFunction}. Our result is plotted in Fig.~\ref{fig:TEMT}. From this figure, one can easily see that the TEMT becomes a density independent nonzero constant after $n_{\rm DLFP}$. This means that, the conformal symmetry is NOT restored after $n_{\rm DLFP}$. Using the relation between TEMT $\theta_\mu^\mu$ and the speed of sound
\be
\frac{\partial \theta_\mu^\mu}{\partial n} = \frac{\partial \epsilon(n)}{\partial n} (1-3 v_s^2),
\ee
one can see that, since $\theta_\mu^\mu$ is independent of density, the left hand side of this equation is zero. Because $\epsilon(n)$ varies with density, $\frac{\partial \epsilon(n)}{\partial n} \neq 0$, one has $1-3 v_s^2 = 0$, i.e., $v_s^2 = 1/3$. Since the TEMT is not zero after $n_{\rm DLFP}$ but the speed of sound satisfies the conformal limit, the matter is pseudoconformal. Therefore, one can conclude neither the emergent of the conformal symmetry nor an onset of the deconfined quark matter from the conformal speed of sound and unit of polytropic index.

\subsection{Minimal model vs Seiberg duality}

What discussed above are based on the minimal model with parameter choice $c_1=-c_2=2/3$ and $c_3=0$~\cite{Meissner:1986js}. So far, no confidential phenomena can uniquely fix these parameters. Here, to show that the existence of DL fixed point is independent of the choice of $c_i$, we choose another set of parameter $c_1=2/3, c_2=-1/3$ and $c_3=1$ ~\cite{Karasik:2020zyo}(Karasik model) which is consistent with the vector meson dominance and the vector meson-gluon duality that realizes the Seiberg duality in the strong interaction sector~\cite{Seiberg:1994pq}.

\begin{figure}[htp]
	\includegraphics[width=9cm]{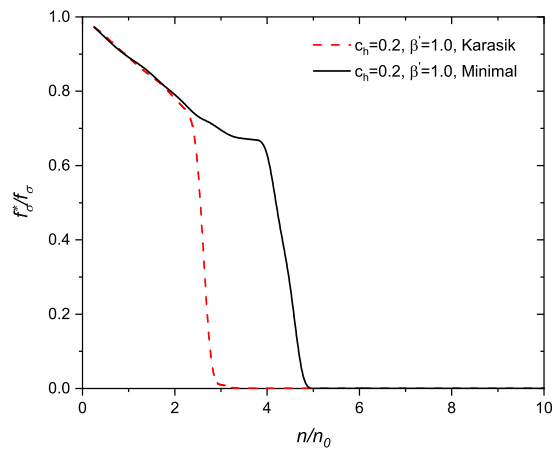}
	\caption{Medium modified dilaton decay constant $f_\sigma^\ast$ from the minimal model (solid) and Karasik model (dashed).}
	\label{fig:fsigmaKM}
\end{figure}

We plot $f_\sigma^\ast$ from the minimal model and the Karasik model in Fig.~\ref{fig:fsigmaKM} for $c_h = 0.2$ and $\beta^\prime = 1.0$. One can see that, the value of $n_{\rm DLFP}$ is smaller in the Karasik model than that in the minimal model. This is because, compared to the minimal model, there is an extra attractive interaction from rho meson in the Karasik model. Effectively, the role of this extra rho meson interaction is identical to increase $\beta^\prime$ in the minimal model.

More importantly, this comparison indicates that, in a dilaton compensated Skyrme type model with a proper choice of the parameters, the existence of DL and VM fixed points is a genuine conclusion.

\subsection{The cores of massive stars}

We finally confront the EoS calculated in the present work with that constrained from astrophysics including the gravitational wave detection. The purpose is to simply give a qualitative idea since only in the large $N_c$ limit a skyrmion can be regarded as a baryon. The constraint (green area) is from~\cite{Annala:2017llu} with upper bound of neutron star $\gtrsim 2.0 M_\odot$. Since the EoS from the crystal approach has negative pressure at $n \lesssim 5.7 n_0$, we only consider the EoS in the region $5.7n_0 \lesssim n \lesssim 10n_0$.

\begin{figure}[htp]
\includegraphics[width=8cm, height=6cm]{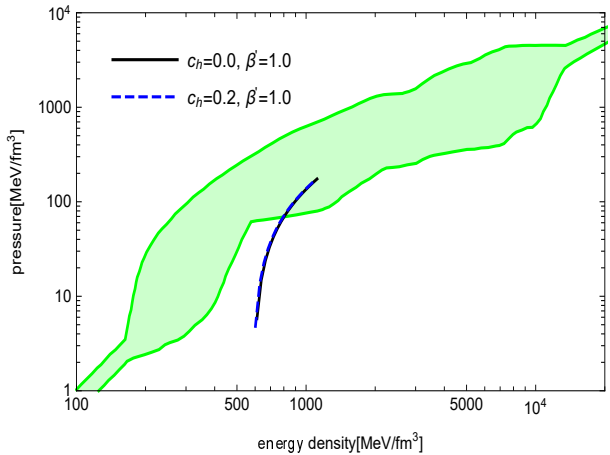}
\caption{Comparison of the EoS calculated in this work upto $10n_0$ and that from the constraint from astrophysics. }
\label{fig:StarCore}
\end{figure}

The comparison of the presently calculated EoS with the constraint is shown in Fig.~\ref{fig:StarCore}. We find that at density $n\gtrsim 7.5 n_0 > n_{\rm DLFP}$ the EoS falls into the region of the constraint. This is not difficult to understand if one respects the fact that a baryon can be regarded as a skyrmion in the sense of large $N_c$ limit and the nuclear matter at low density is not a crystal solid therefore the present numerical results make sense at large density.


\section{Conclusion and Discussion}

\label{sec:Conclusion}

We study in this paper the properties of dense nuclear matter using the skyrmion crystal approach based on $s$HLS including the lightest scalar meson $\sigma$, isovector vector meson $\rho$, isoscalar vector meson $\omega$, in addition to the pseudo-Nambu-Goldstone boson $\pi$.

We found that the requirement of the scale symmetry restoration in dense nuclear matter, a natural expectation from physics, constrains the magnitudes of the scale symmetry breaking as $0 \lesssim c_h \lesssim 0.2$ and $1.0 \lesssim \beta^\prime \lesssim 2.0$. We also found that after $n_{\rm DLFP}$, both the speed of sound and the polytropic index approach to the conformal limit $v_s^2 \to 1/3$ and $\gamma \to 1$ while the TEMT does not vanish but becomes a density independent constant. Therefore, the dense nuclear matter is a pseudoconformal one. To have a qualitative idea, we compare the EoS calculated in the present approach and the constraint from heavy ion physic and astrophysics.

Since the TEMT is a nonzero constant, our matter is still in the hadronic phase after DL fixed point, not in the quark phase. This means that, in contrast to~\cite{Annala:2019puf}, in nuclear matter, neither the conformal speed of sound nor the smallness of the polytropic index can be used as a sufficient criterion of the onset of the quark matter. The constant TEMT after DL fixed point may closely related to chiral invariant mass of nucleon~\cite{Detar:1988kn} which has been found significant in constructing the EoS of nuclear matter~\cite{Motohiro:2015taa}. The chiral invariant mass of nucleon may attribute to the multiquark condensate and gluon condensate in the present approach.


\acknowledgments

Y.~L.~M would like to thank Mannque Rho for his valuable discussions and comments. The work of Y.~L. M. was supported in part by National Science Foundation of China (NSFC) under Grant No. 11875147 and No. 12147103.



\bibliography{reference}


\end{document}